\documentclass[iop]{emulateapj}
\usepackage{graphicx,epsfig,fancyhdr,rotating,amsmath,natbib}
\usepackage{txfonts}
\usepackage{epsfig,rotate}
\usepackage{natbib}

\shorttitle{}
\shortauthors{Sharma et al.}
\begin{document}
\title{Direct observations of different sunspot waves influenced by umbral flashes}
\author{Aishawnnya Sharma\altaffilmark{1,2}, G.~R. Gupta\altaffilmark{2}, Durgesh Tripathi\altaffilmark{2}, V. Kashyap\altaffilmark{3}, \and Amit Pathak\altaffilmark{1,4}}
\affil{\altaffilmark{1}Department of Physics, Tezpur University, Tezpur 784028, India}
\affil{\altaffilmark{2}Inter-University Centre for Astronomy and Astrophysics, Post Bag-4, Ganeshkhind, Pune 411007, India}
\affil{\altaffilmark{3}Harvard-Smithsonian Center for Astrophysics, Cambridge, MA, USA}
\affil{\altaffilmark{4}Current Address: Department of Physics, Banaras Hindu University, Varanasi 221005, India}
\email{girjesh@iucaa.in}

\begin{abstract}
We report the simultaneous presence of chromospheric umbral flashes and associated umbral waves, and propagating coronal disturbances, in a sunspot and related active region. 
We have analyzed time-distance maps obtained using the observations from Atmospheric Imaging Assembly (AIA) on-board Solar Dynamics Observatory (SDO). These maps show
the simultaneous occurrence of different sunspot oscillation and waves such as umbral flashes, umbral waves, and coronal waves. Analysis of the original light curves,
i.e., without implementing any Fourier filtering on them, show that the amplitudes of different sunspot waves observed at different atmospheric layers change in synchronization
with the light curves obtained from the umbral flash region, thus demonstrating that these oscillations are modulated by umbral flashes. This study provides the first observational 
evidence of the influence of sunspot oscillations within the umbra on other sunspot waves extending up to the corona. The properties of these waves and oscillations can be utilized
to study the inherent magnetic coupling among different layers of the solar atmosphere above sunspots.
\end{abstract}


\section{Introduction}\label{intro}

Waves  play an important role in the heating of upper atmosphere of the Sun. Different features observed over 
sunspots at different atmospheric heights host a variety of waves, such as the 5-min photospheric oscillations, 
the 3-min chromospheric oscillations, umbral flashes and waves, running penumbral waves, and propagating coronal waves
\citep[see for e.g., reviews by][]{2006RSPTA.364..313B,2012RSPTA.370.3193D,
2016GMS...216..467S}. Although these oscillations and waves have been studied for decades, we are still far from 
understanding the physics behind their origin and the possible coupling among them. It has further been suggested that sunspot waves and oscillations 
may play an important role in the initiation of solar flares and coronal mass ejections (CMEs), 
as well as solar wind acceleration \citep[see, e.g.,][]{2016GMS...216..467S}. Recent studies show that sunspot 
waves may also play an important role in the triggering of coronal jets \citep{2015MNRAS.446.3741C}.
Jets were triggered during the growing amplitude phase of the waves, however,
the cause of such an amplitude increase is still unknown.

Umbral flashes are observed as sudden strong brightenings occurring at random locations in the sunspot umbrae with 
a period of around 3-min in chromospheric lines and are considered as the first observations of sunspot oscillations 
\citep{1969SoPh....7..351B}. These are strongly non-linear oscillations with asymmetric light curves, where the increase 
in the amplitude is steeper than the decrease,  giving it a saw-tooth shape. Such light curves are interpreted as signatures of upward propagating 
magneto-acoustic shock waves \citep[e.g.][]{2003A&A...403..277R,2006ApJ...640.1153C}. The shock wave
nature of sunspot oscillations has also been recently reported in the transition region lines \citep{2014ApJ...786..137T}. The running 
penumbral waves (RPW) are outward propagating intensity waves with a period of about 5-min and are observed in chromospheric 
penumbrae of sunspots \citep{1972ApJ...178L..85Z}. These oscillations are interpreted as upward propagating
magneto-acoustic waves guided by the magnetic field and originate in the lower atmosphere \citep{2007ApJ...671.1005B,
2013ApJ...779..168J}.

\begin{figure*}[htbp]
\centering
\includegraphics[width=0.9\textwidth]{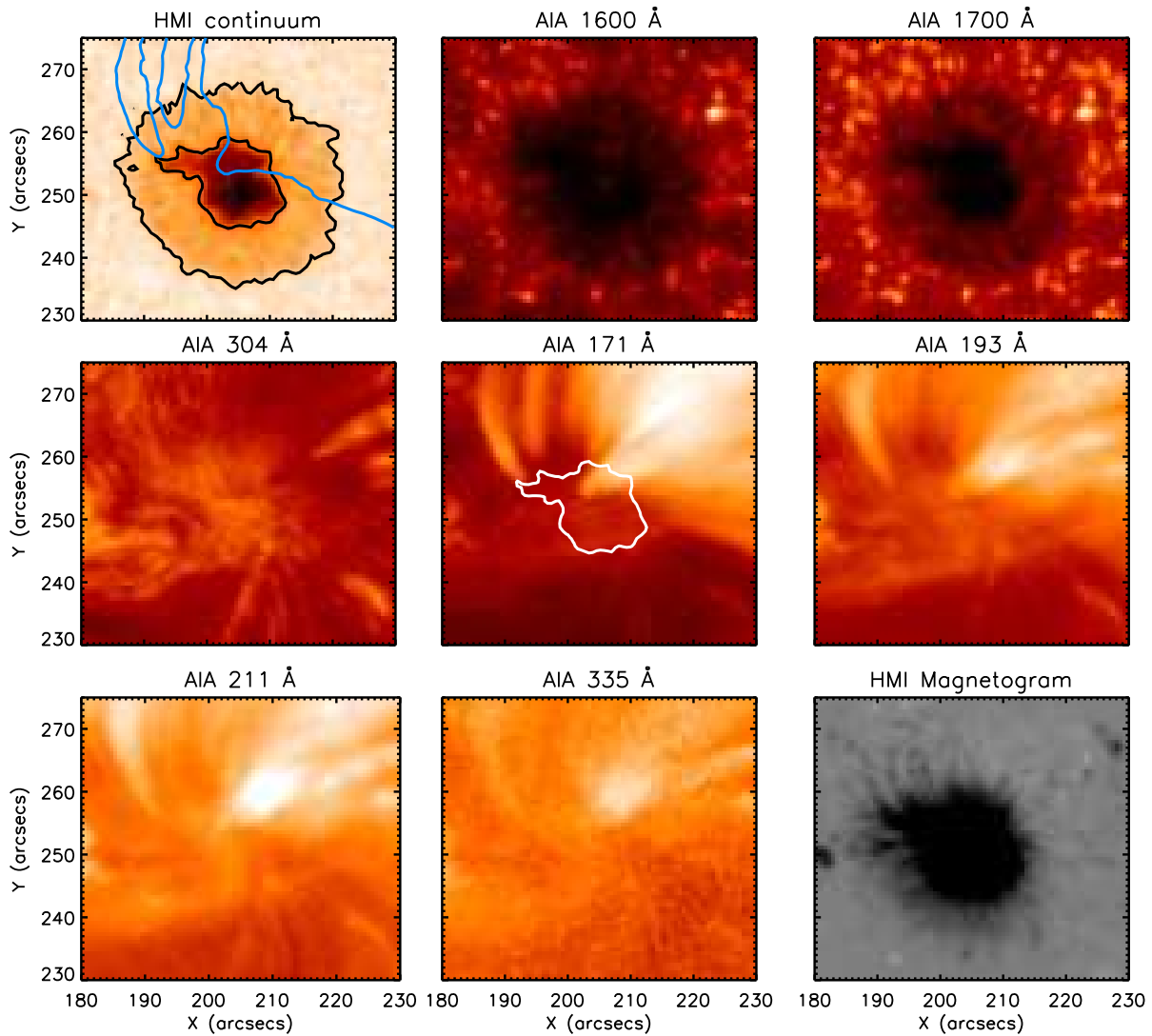}
\caption{Analyzed sunspot region observed in different AIA and HMI filters. Inner black contour on 
the top of HMI continuum image shows the boundary between umbra and penumbra, whereas outer one shows 
penumbra outer boundary obtained from HMI continuum.The umbra-penumbra boundary
is also shown by a white contour in AIA 171~{\AA} for reference. Over-plotted blue contours on HMI continuum
show the locations of fanloops observed in AIA 171~{\AA} passband. }
\label{fig:sunspot}
\end{figure*}

The relationship between 3-min umbral waves and 5-min running penumbral waves are still not fully understood.
While some studies have advocated that they are the different manifestations 
of a common phenomenon \citep{2001A&A...375..617C,2006A&A...456..689T,2008sust.book.....T}, other studies suggest an unclear relationship 
between them \citep{2000A&A...354..305C,2004A&A...424..671K,2006SoPh..238..231K}.   
Recently \citet{2015ApJ...800..129M} have claimed that both umbral flashes and running waves originate from photospheric p-mode
oscillations, where umbral flashes were preceding the running waves in both the spatial and temporal domains.

Propagating intensity disturbances along various coronal structures with the period between 3-20 min are  ubiquitous in
the solar corona \citep{2009SSRv..149...65D,2012RSPTA.370.3193D}. The loop like structures, which are often rooted in the umbra 
show outward propagating intensity disturbances with periods around 3-min, whereas those rooted in non-sunspot regions show 
periods around 5-min \citep[e.g.][]{2002A&A...387L..13D}. Furthermore, open plume and interplume structures in the polar region
also show outward propagating
intensity disturbances with periods around 10--30 min  \citep[e.g.][]{2010ApJ...718...11G,2011A&A...528L...4K}. These propagating 
disturbances are found to have wave-like properties and are often interpreted in terms of propagating slow  magneto-acoustic waves
\citep[e.g.][]{2012SoPh..279..427K,2012A&A...546A..93G}. Although these coronal wave disturbances are ubiquitous in the different 
structures, observational evidence of their source region is still missing.  

Recently, \citet{2012ApJ...757..160J} found 3-min magneto-acoustic waves in the coronal fanloops which were rooted into 
the photosphere at locations where large-amplitude 3-min umbral dot oscillations were observed. 
\citet{2015ApJ...812L..15K} compared the period of amplitude modulation on Fourier-filtered light curves obtained in
different atmospheric layers above the sunspot and associated the presence of  slow magneto-acoustic waves in coronal 
loops with the photospheric p-mode.  \citet{2016ApJ...830L..17Z} traced p-mode waves from the photosphere to the 
corona in active regions using a time-distance helioseismology analysis technique. 
However, direct observation of any connection or influence among different sunspot waves and
oscillations at different atmospheric layers is still missing. 

For direct and unambiguous detection, it is mandatory to have excellent wave signal at different  atmospheric 
layers, which is not always the case. Here, we present an observation where sunspot oscillations
were strong enough to show the influence of the perturbation caused by one of the waves on the other waves.
Previous such analyses have utilized light curves at individual locations at  different atmospheric heights
and performed co-spatial analysis. However, here we present multi-wavelength analysis on various locations obtained 
from the time-distance plots. It has helped us to establish a connection between waves in different layers of the solar atmosphere
using observations recorded by the Atmospheric Imaging Assembly \citep[AIA,][]{2012SoPh..275...17L} on-board
the Solar Dynamics Observatory \citep[SDO,][]{2012SoPh..275....3P}. We show that umbral flashes influence 
the propagation of umbral and coronal waves and investigate the characteristics of the different
waves with respect to each other. We present the details of the observations in \S~\ref{obs}, data analysis and
results in \S~\ref{analysis}, and finally summarize our results and conclude in \S~\ref{conclusion}.

\begin{figure*}[htbp]
\centering
\includegraphics[width=0.98\textwidth]{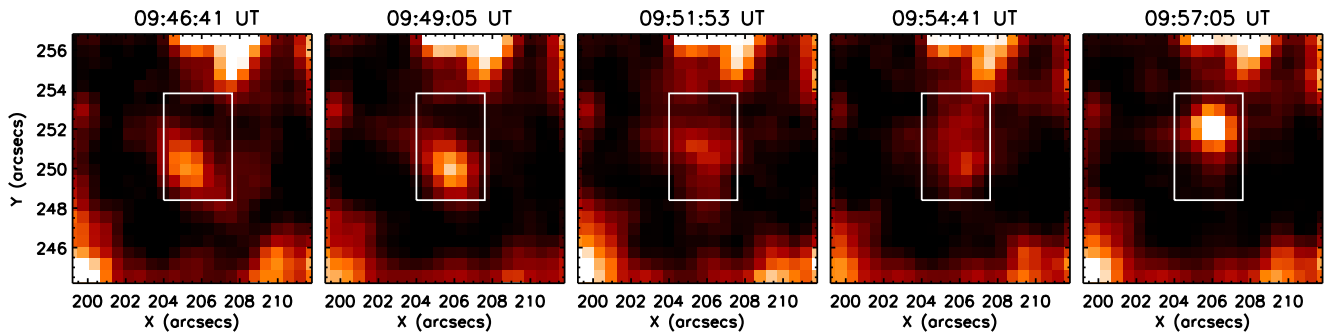}
\caption{Five umbral flashes observed in AR 11133 in the AIA 1600~{\AA} passband. The white 
boxes indicate the region within which the flashes occured.}
\label{fig:uf}
\end{figure*}

\begin{figure*}[htbp]
\centering
\includegraphics[width=0.8\textwidth]{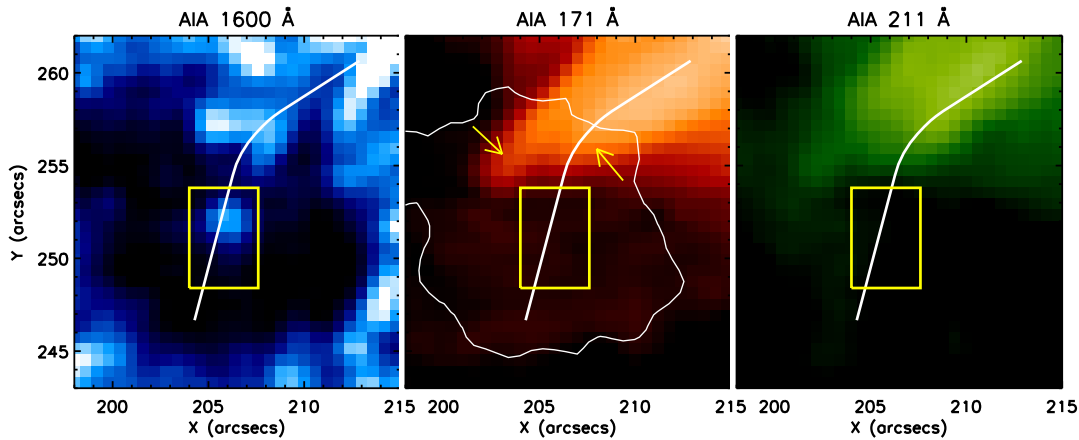}
\caption{Location of the artificial slit drawn on the top 
of AIA 1600 \AA\ (left panel), 171 \AA\ (middle panel), and 211 \AA\ (right panel) images along which a 
time-distance analysis is performed. Yellow boxes on the images show the location of 
the flashes observed in the AIA 1600 \AA\ passband (shown in Figure~\ref{fig:uf}). White contour on the top of
AIA 171 \AA\ marks the location of approximate umbra-penumbra boundary obtained from HMI continuum. Overplotted yellow arrows are directed to
the two fanloop systems rooted inside sunspot umbra.}
\label{fig:slit}
\end{figure*}

\section{Observations}\label{obs}

We have analyzed the multi-wavelength observations of an active region (AR) NOAA \textit{AR 11133} observed by SDO
on December 11, 2010 between 09:30:00 to 10:15:00~UT. We have used AIA/SDO observations in two of its UV channels 
(1700~{\AA}, and 1600~{\AA}) and all of its EUV channels (304~{\AA}, 131~{\AA}, 171~{\AA}, 193~{\AA}, 211~{\AA}, 
335~{\AA}, and 94~{\AA}). The datasets for UV have a cadence of 24~s, while those of the EUV channels have a cadence 
of 12~s. We have also used data from Helioseismic and Magnetic Imager (HMI) on-board SDO to provide context.
The cadence of HMI data is 45~s. The spatial resolution of both AIA and HMI images are $0.6\arcsec$ per pixel. 
The AIA and HMI observations are processed using standard processing software 
provided in the solar software (SSW) distribution. 
All the images are co-aligned and de-rotated with respect to the AIA 171~{\AA} image taken at 9:30:00 UT. 

\section{Data Analysis and Results} \label{analysis}

The observed AR mainly consists of a sunspot with fanloops emanating from its upper half. Figure~\ref{fig:sunspot} 
displays the AR in different AIA and HMI passbands. The top left panel shows the analyzed active region
in HMI continuum. Black contours obtained from the 
HMI continuum show the approximate locations of  umbra-penumbra (inner contour) and 
penumbra outer (outer contour) boundaries. Over-plotted blue contours show the fanloop configuration as observed in 
the AIA 171~{\AA} passband.  

\subsection{Umbral Flashes, Umbral Waves, and Coronal Waves}

\begin{figure*}[htbp]
\centering
\includegraphics[width=0.9\textwidth]{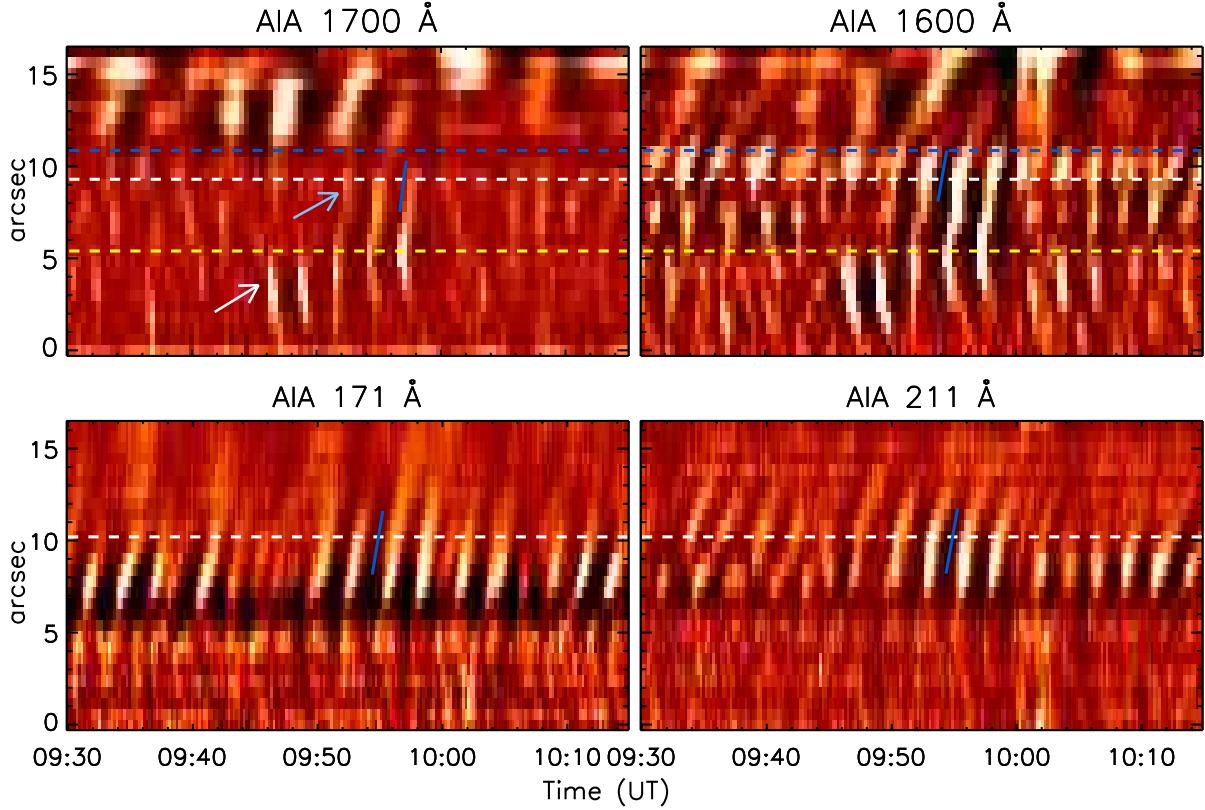}
\caption{Time-distance plots obtained from different AIA passbands along the artificial slit location shown in Figure~\ref{fig:slit}. 
In the top left panel, the white arrow points to umbral flashes and the blue arrow points to umbral waves.
Dashed yellow horizontal lines indicate the location of umbral flashes. White horizontal lines
on each panel show the locations of light curves obtained for further
analysis. Blue horizontal lines indicate the umbra-penumbra boundary identified from HMI continuum.
Slanted blue lines along propagating features in each panel are used to measure the average wave propagation speed.}
\label{fig:xt}
\end{figure*}

\begin{figure}[htbp]
\centering
\includegraphics[width=0.46\textwidth]{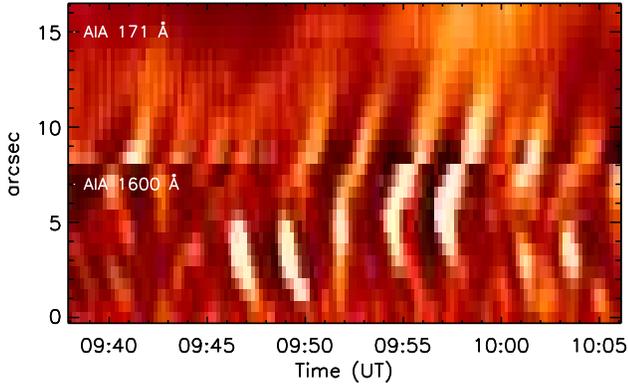}
\caption{Combined time-distance plot obtained from AIA 1600 \AA\ and 171 \AA\ passbands for the artificial slit location shown in Figure~\ref{fig:slit}.
AIA 1600 \AA\ is plotted from $0\arcsec$\ to $8\arcsec$\ whereas AIA 171 \AA\ is plotted from $8\arcsec$\ to $16\arcsec$.  Intensities are normalized 
by time averaged variation along the slit length.} 
\label{fig:mxt}
\end{figure}

We spotted five bright umbral flashes between 09:46:41 UT and 09:57:05 UT in AIA 1700~{\AA} and 1600~{\AA} passbands,
as shown in Figure~\ref{fig:uf}. The over-plotted white-box encloses the region within which the different umbral flashes occur.
In Figure~\ref{fig:slit}, we show the location of umbral flashes on AIA 1600~{\AA}, 171 \AA , and 211 \AA\ images.
We overplot the approximate umbra-penumbra boundary (white contour) obtained from HMI continuum on the
top of AIA 171~{\AA} image (middle panel of Figure~\ref{fig:slit}). We find that there are two fanloop systems
with their coronal footpoints located at different locations of the sunspot umbra
(as marked by yellow arrows in Figure~\ref{fig:slit}).
To study the effect of perturbation caused by umbral flashes on the surrounding 
sunspot waves, we adopt a time-distance analysis technique. We show the location of the artificial slit to be used
for time-distance technique in Figure~\ref{fig:slit}. We choose the artificial slit in such a way that it passes through 
the umbral flashes and also traces a fanloop to observe any influence of flashes on the fanloop.
In Figure~\ref{fig:xt}, we show the time-distance maps obtained along this slit in different AIA passbands 
covering the chromosphere and corona above the sunspot. Maps were obtained by subtracting the
background trend of $\approx 8$-min running average from each spatial pixel along the time.
We tried several ranges of running average windows, and found that 8-min running averages represent the background/trend signal very well.
The time-distance maps clearly show the presence of propagating disturbances in the different layers of the sunspot atmosphere.
Five umbral flashes at chromospheric height can be seen in the  upper panels of AIA 1600 \AA\ and 1700 \AA .
The  white arrow in AIA 1700 \AA\ panel  locates the umbral flashes. 
In the top panels of Figure~\ref{fig:xt}, the yellow dashed lines pass through the approximate location of
the umbral flashes, whereas, the white dashed lines pass through the umbral waves. The blue dashed lines show the umbra-penumbra boundary. 
The time-distance maps clearly reveal the presence of umbral waves emanating from the location of umbral flashes and moving radially outward.
Umbral waves are found to be confined to the region between the location of umbral flashes and the umbra-penumbra boundary i.e.,
the region between blue and yellow dashed lines in Figure~\ref{fig:xt}. The blue arrow in the AIA 1700 \AA\  panel shows the propagation of 
umbral waves originating from the location of umbral flashes. We drew several lines on these propagating features and obtained the average 
slope and standard deviation which provided the wave propagation speed and associated errors. 
The umbral wave speeds are found to be quite similar (within errors) in different passbands  with around
$66.1\pm 8.7$~km~s$^{-1}$ for 1700~{\AA}, $49.0\pm 7.1 $~km~s$^{-1}$ for 1600~{\AA}, and $56.7\pm 5.1$~km~s$^{-1}$ for 304~{\AA} passbands.

Propagating coronal waves are omnipresent along the fanloop in all the AIA coronal passbands for the observed time duration 
except in 94~{\AA}, where the signal is too poor to make any conclusive statement. 
Coronal waves are also detectable for the other fanloops of umbral and penumbral origin
(i.e., coronal footpoints co-spatial to umbra and penumbra of the sunspot) as visible 
in the coronal images of Figure~\ref{fig:sunspot}. 
In the bottom panels of Figure~\ref{fig:xt}, we show the presence of coronal waves for AIA 171, and AIA 211~{\AA} passbands propagating
along the analyzed fanloop rooted in the umbra. The white dashed lines in the bottom panels of Figure~\ref{fig:xt} pass through the coronal waves.
The coronal wave speeds  are found to be around $50.9\pm 4.9$~km~s$^{-1}$ for 171~{\AA},
$46.2\pm 5.3$~km~s$^{-1}$ for 193~{\AA}, $46.9\pm 3.6$~km~s$^{-1}$ for 211~{\AA}, $62.4\pm 9.2$~km~s$^{-1}$ for 335~{\AA},
and $44.8\pm 6.2$~km~s$^{-1}$ for 131~{\AA} passbands. 

\begin{figure*}[htbp]
\centering 
\includegraphics[width=7.cm,angle=90]{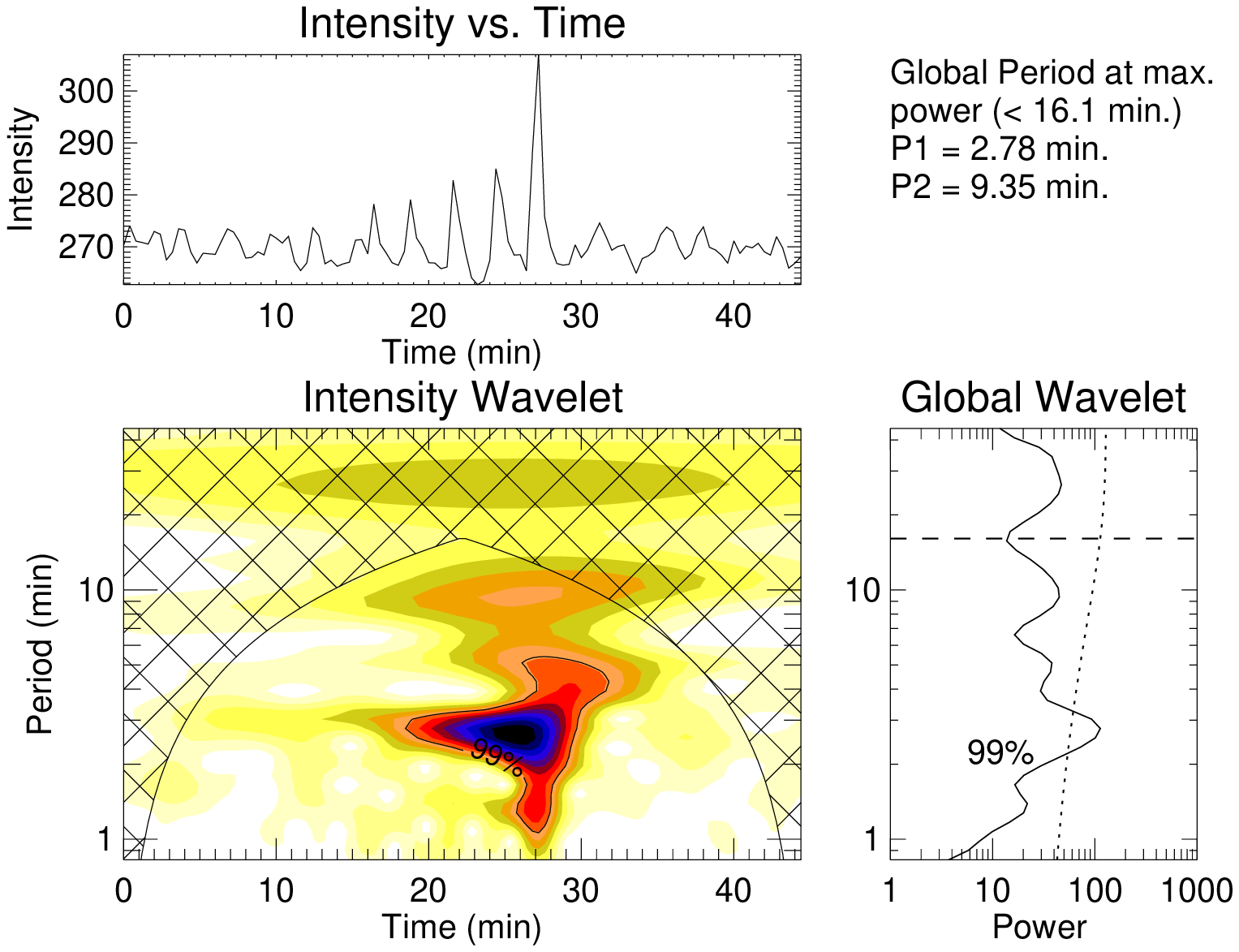}\includegraphics[width=7.cm,angle=90]{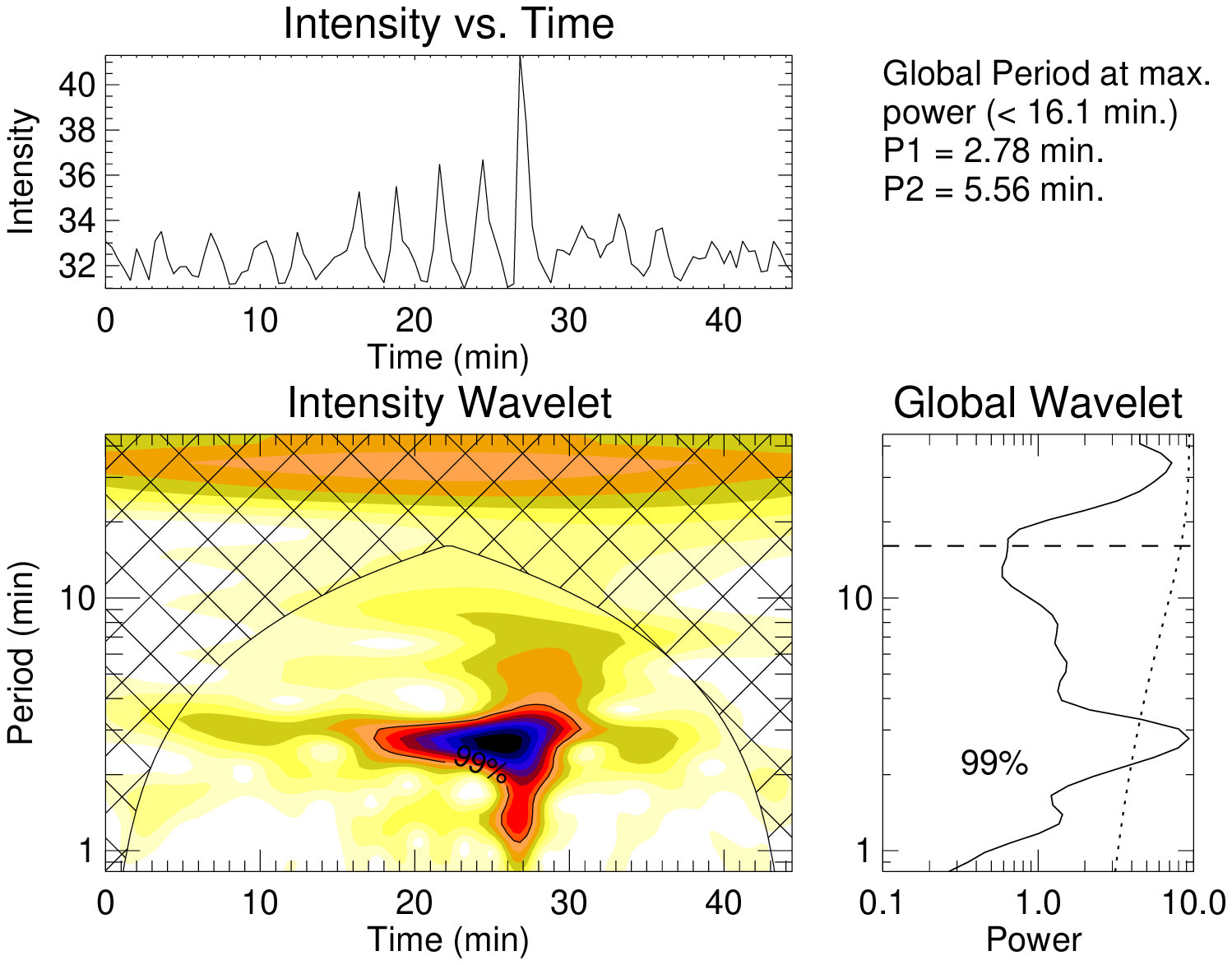}
\caption{Wavelet analysis results for the light curves obtained at the umbral flash location (shown in Figure~\ref{fig:xt}) observed in 
AIA 1700 \AA\ (left panels) and AIA  1600~{\AA} (right panels) passbands. In each set, the top panels show the variation of measured 
intensity with time where time starts around 9:30 UT. The bottom left panels show the computed wavelet power 
spectrum (blue shaded represents high power density), while the bottom right panels 
show the global wavelet power spectrum. Dashed lines in the global wavelet plots indicate the
maximum period detectable from wavelet analysis due to cone-of-influence whereas the dotted line indicates 99\% confidence level curve.
Periods P1 and P2 of the first two power peaks are also printed at the top right.}
\label{fig:ufwavelet}
\end{figure*}

\begin{figure*}[htbp]
\centering
\includegraphics[width=7.cm,angle=90]{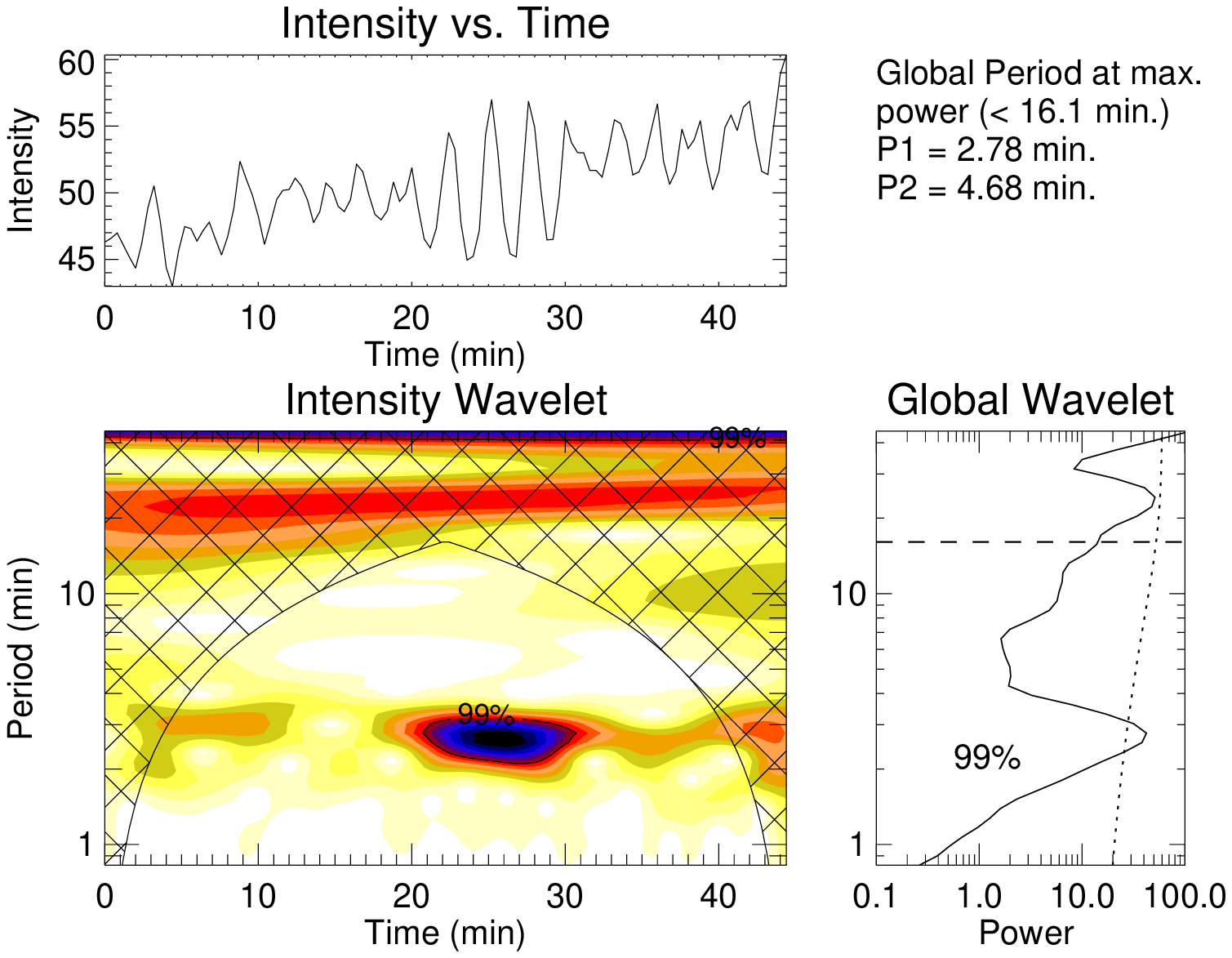}\includegraphics[width=7.cm,angle=90]{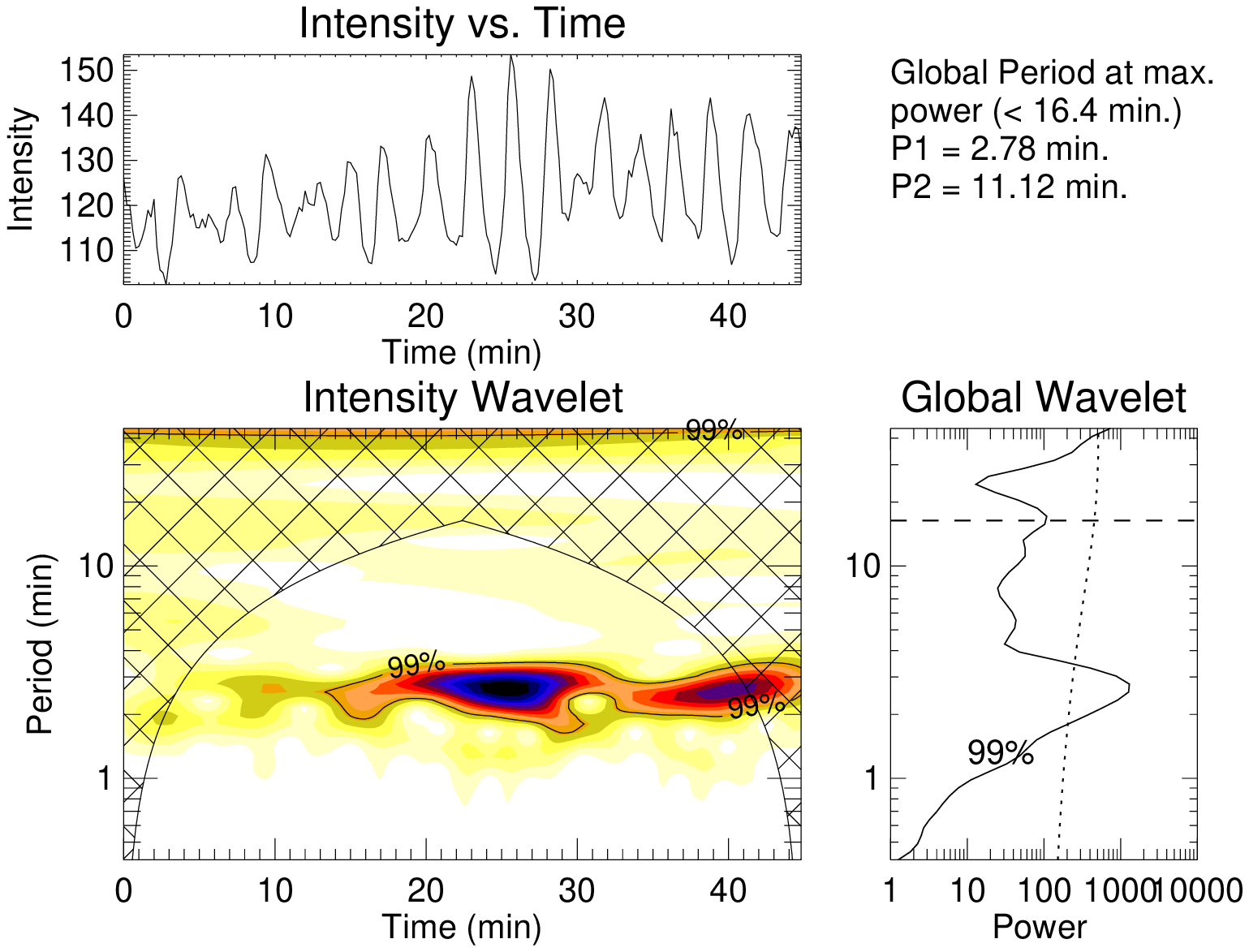}
\caption{Same as Figure~\ref{fig:ufwavelet}, but for umbral waves observed in AIA 1600~{\AA} (left panels) and AIA 304~{\AA} (right panels).}
\label{fig:uwwavelet}
\end{figure*}

The time-distance maps reveal a peculiar noticeable characteristic for the different sunspot waves. 
We find an enhancement in the amplitude of the umbral and the coronal waves for the duration of  occurrence 
of the five bright umbral flashes. Enhancements in the amplitude of coronal waves, which resulted in the triggering of coronal jets
were also observed by \citet{2015MNRAS.446.3741C}. In order to have a clear picture of
the simultaneous amplitude enhancement between different sunspot oscillation and wave modes, 
 we show a combined time-distance map of chromospheric AIA 1600~{\AA}
and coronal AIA 171~{\AA} passbands in Figure~\ref{fig:mxt}. Cadence of AIA 1600 \AA\ images is 24 s whereas that of AIA 171 \AA\
images is 12 s. Therefore, we interpolated the AIA 1600 \AA\ images to 12 s cadence to create the combined time-distance map. 
In this map, we plot AIA 1600 \AA\ from $0\arcsec$\ to $8\arcsec$\ and AIA 171 \AA\ from $8\arcsec$\ to $16\arcsec$. 
The resulting map clearly shows an amplitude increase in coronal waves associated with the occurrence of umbral flashes, and
thus, with umbral waves. The time delay between the two is about 36 s (3-time frames of AIA 171 \AA ). This indicates that umbral flashes influence the
propagation of coronal waves, providing us with the first direct evidence of an influence of umbral flashes on the coronal plasma.
We also analyzed the propagation of coronal waves in other fanloops of umbral and penumbral origin,
rooted in the same sunspot. In this case, we did not find any influence of  umbral flashes 
in terms of amplitude enhancement in coronal waves propagating along the fanloops of penumbral origin.
However, coronal waves of other fanloop system rooted in the umbra (left loop in Figure~\ref{fig:slit}) did 
show some influence of umbral flashes. 

The time-distance maps obtained along the artificial slit suggest a growth in 
the amplitude of the waves during 09:44 to 10:00 UT. To analyze this in detail, we
obtain light curves at the locations of the umbral flash (yellow dashed line in Figure~\ref{fig:xt}) and 
the umbral and coronal waves (white dashed lines in Figure~\ref{fig:xt}). We choose the locations on 
the basis of signal strength. The umbral flash location is averaged
over $3\arcsec$, while the umbral and coronal wave locations are averaged over $1.2\arcsec$\ and $1.8\arcsec$\ respectively.
The detailed analysis performed on these light curves are described in the following subsections.

\subsection{Wavelet Analysis}

\begin{figure*}[htbp]
\centering
\includegraphics[width=7cm,angle=90]{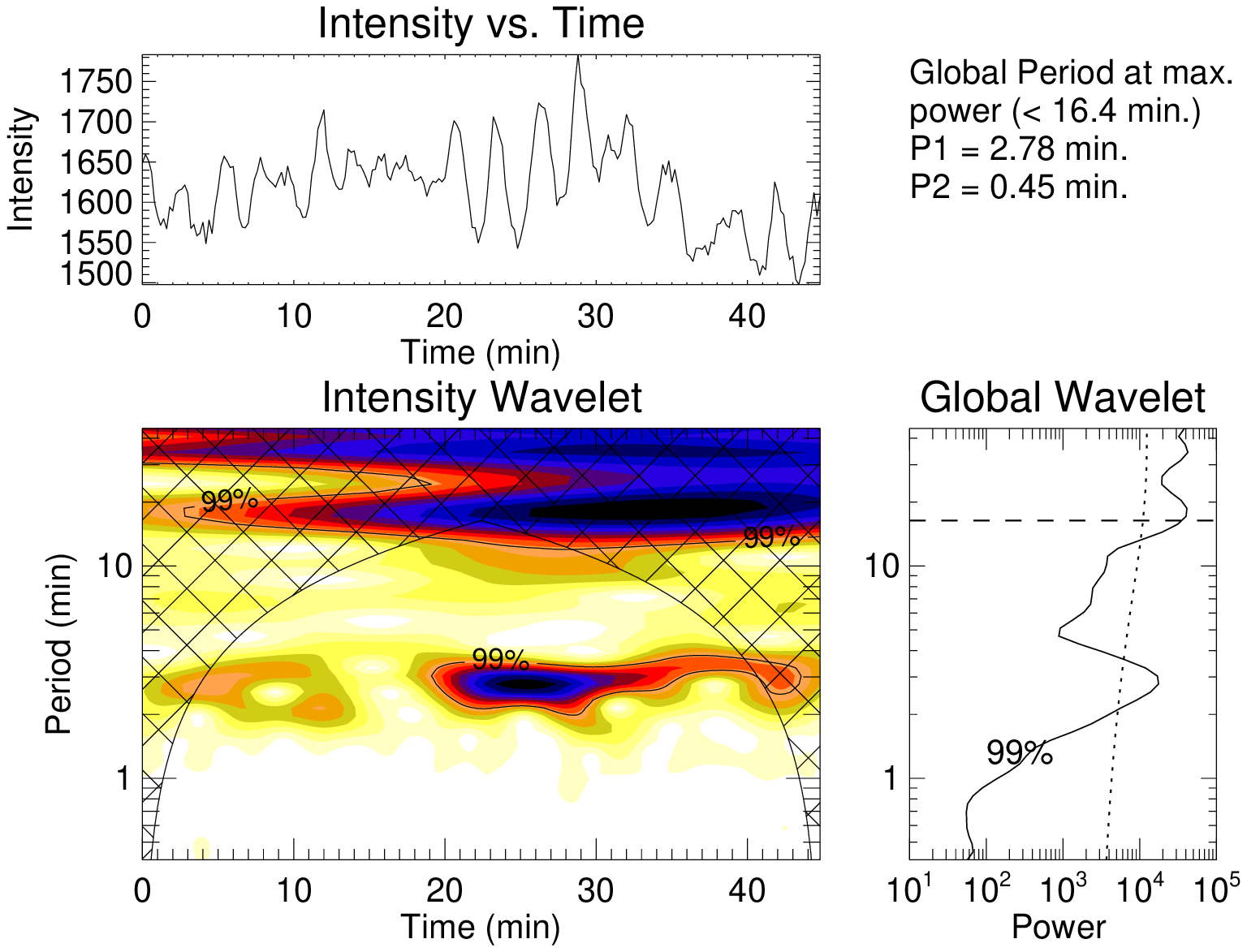}\includegraphics[width=7cm,angle=90]{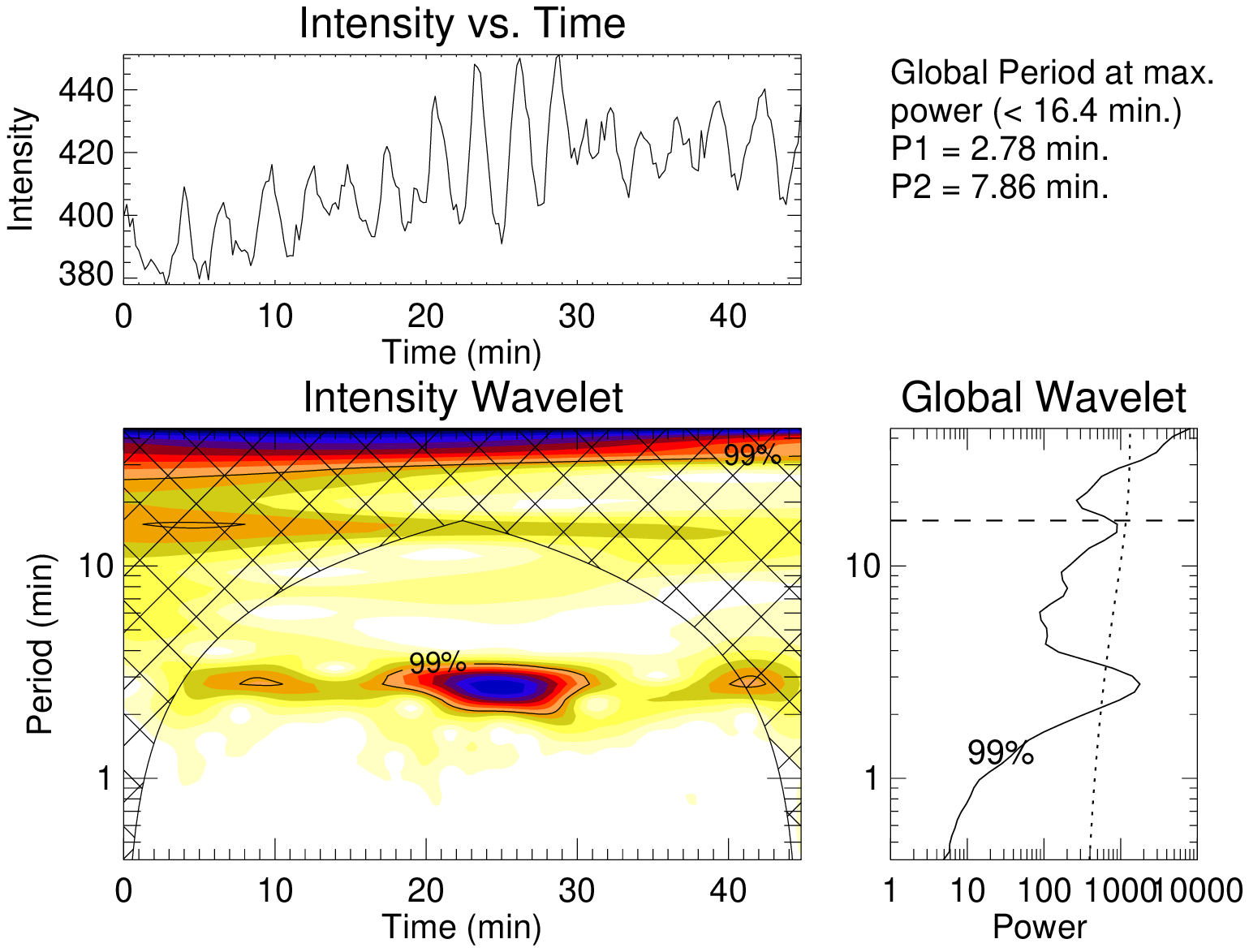}
\caption{Same as Figure~\ref{fig:ufwavelet}, but for coronal waves observed in AIA 171~{\AA} (left panels) and AIA 211~{\AA} (right panels).}
\label{fig:cwwavelet}
\end{figure*}

We obtain temporal intensity variations of the umbral flash region, umbral waves, and coronal waves for locations 
marked in Figure~\ref{fig:xt}. The time evolution of intensities obtained from various AIA passbands for different sunspot waves 
are plotted in the top panels of Figures~\ref{fig:ufwavelet}, \ref{fig:uwwavelet}, and \ref{fig:cwwavelet}. All these intensity light curves
show prominent growth in the amplitude of oscillations for the similar time as that of the occurrence of umbral flashes. 
In all figures, time runs from 9:30 UT to 10:15 UT.


\begin{figure*}[htbp]
\centering
\includegraphics[width=0.72\textwidth]{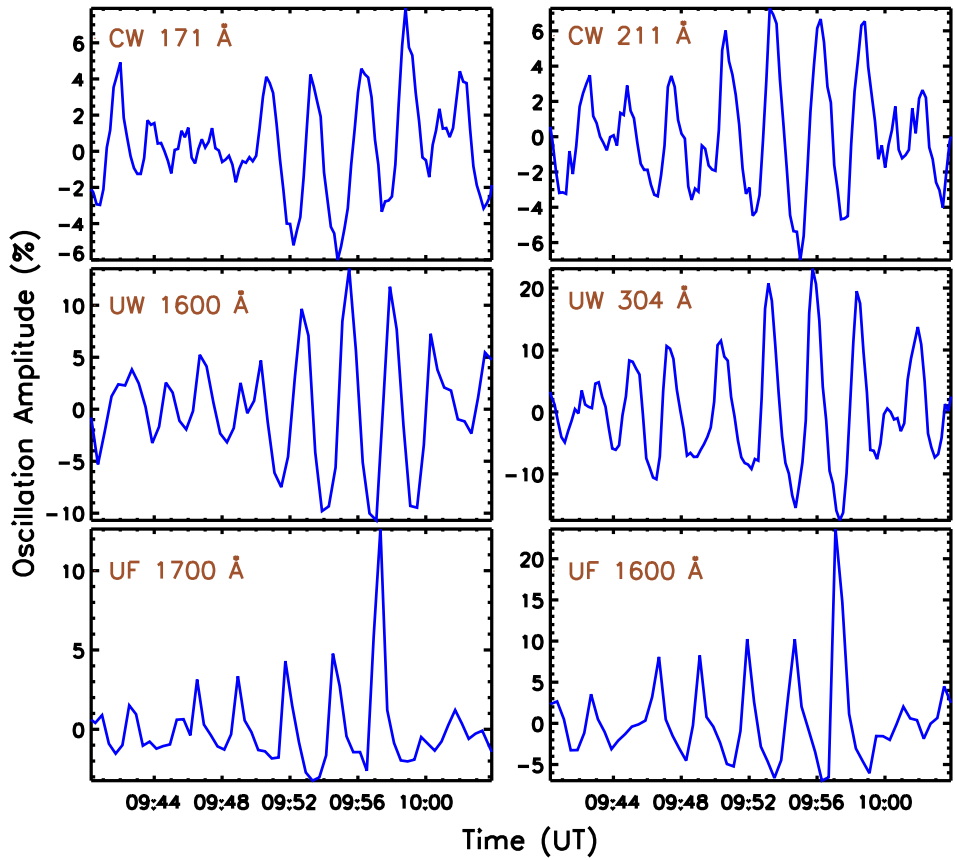}
\caption{Growing amplitude of different type of sunspot waves and oscillations observed at different
layers of the solar atmosphere. The respective locations of light curves used to obtain oscillation
amplitudes are marked in Figure~\ref{fig:xt}.}
\label{fig:growing}
\end{figure*}

\begin{figure*}[htbp]
\centering
\includegraphics[width=0.8\textwidth]{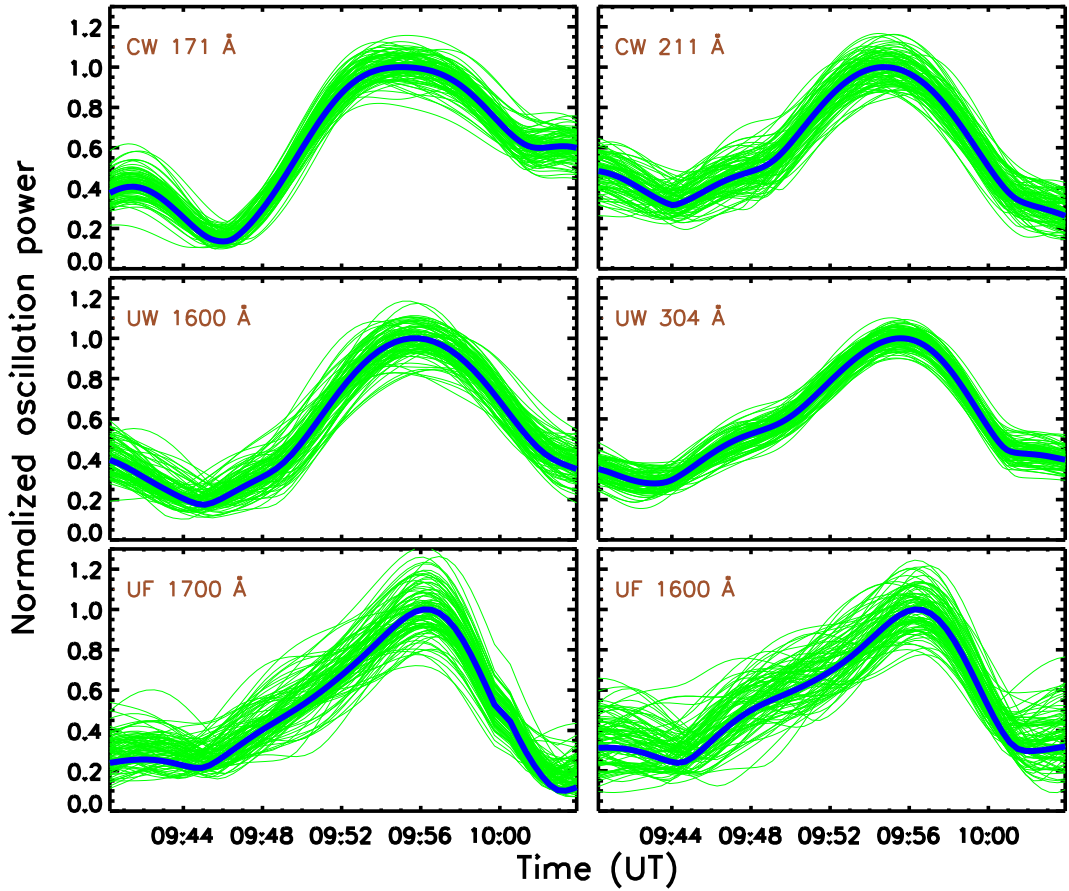}
\caption{Temporal variation of oscillating power in the period range 2.3--3.3 min obtained from wavelet transform of different sunspot 
waves and oscillations. Green curves are obtained by carrying out wavelet analysis on Monte Carlo bootstrapped light curves.
Locations of analysis are marked in Figure~\ref{fig:xt}.}
\label{fig:power}
\end{figure*}

To obtain the period of these oscillations, we performed wavelet analysis \citep{1998BAMS...79...61T} on all the light curves.
Wavelet transform  provides information on the temporal variation of frequency of a signal. 
For this purpose, we chose the Morlet wavelet that is a plane wave with its amplitude modulated by a Gaussian function
to convolve with the time series. In Figures~\ref{fig:ufwavelet}, \ref{fig:uwwavelet} and \ref{fig:cwwavelet}, we show the wavelet 
results for umbral flashes, umbral waves, and coronal waves respectively in different AIA passbands as mentioned in the captions.
In each wavelet spectrum (lower left panels), the cross-hatched regions denote
the so called cone-of-influence (COI) locations where estimates of oscillation periods become unreliable. 
This COI is result of edge effects which arises due to the finite-length of time series.
The global wavelet power, obtained by taking the average over the time domain of the wavelet transform is also shown for all the sets in the lower right panels.
Due to the COI, the maximum period which can be detected from the wavelet transform is shown by a horizontal dashed line in the global 
wavelet plots of Figures~\ref{fig:ufwavelet}, \ref{fig:uwwavelet}, and \ref{fig:cwwavelet}. 
The confidence level of  99\% is  shown in global wavelet plots which are obtained after considering the white noise in the data. 
We also obtained first two power peaks from the global wavelet which are printed at the right top corner of wavelet plots. 
Global wavelet plots for umbral and coronal waves show very similar nature of power distribution near the peak period of $\approx$ 2.8-min.
Results from wavelet analysis reveal the clear presence of $\approx$ 2.8-min period oscillations for all the three sunspot oscillation
and waves over the whole observed duration. However, we also noticed that wavelet powers for this period are not constant and change with time. 
In the time range between $\approx 15-30$ min, wavelet power increases with time for all the three sunspot oscillation and waves, and later decreases.
This almost co-temporal increase in wavelet power with time in different waves is suggestive of coupling  among them which was also visualized
from the time-distance maps in Figure~\ref{fig:xt}.

\begin{figure*}[htbp]
\centering
\includegraphics[width=0.8\textwidth]{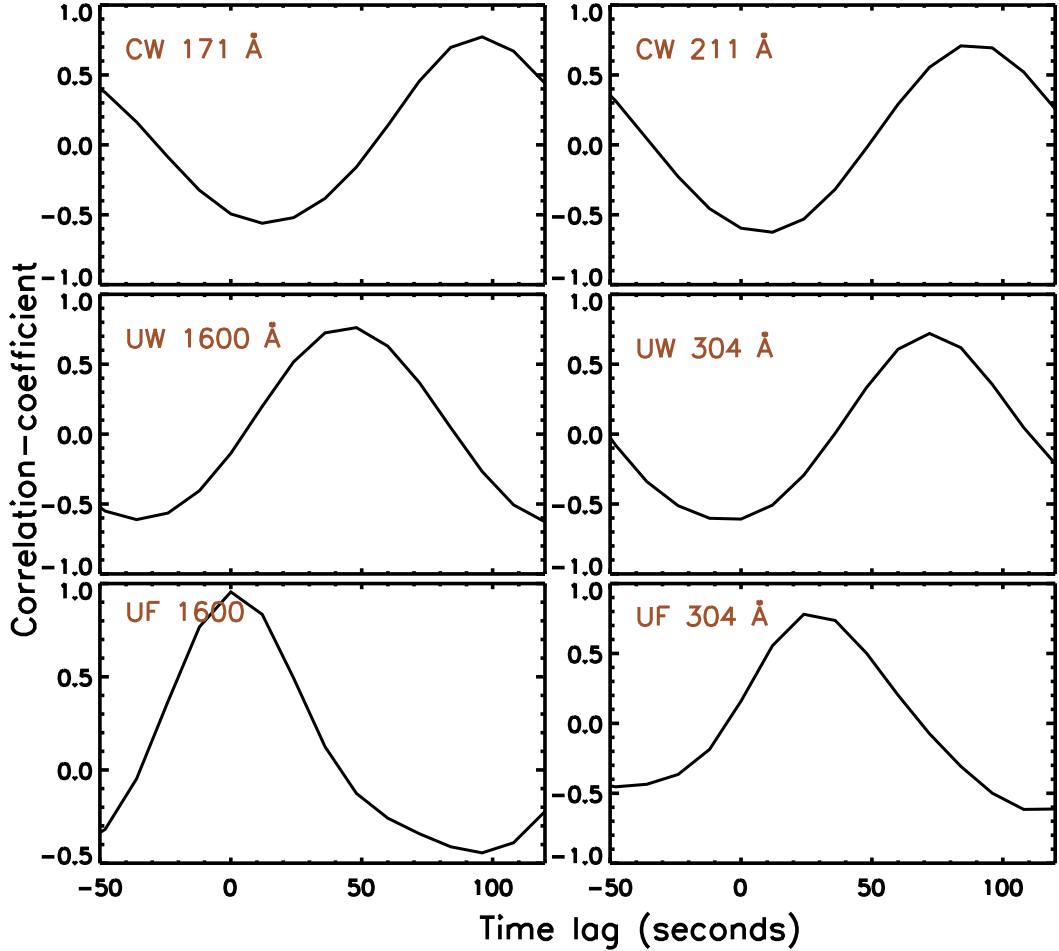}
\caption{Results of cross-correlation analysis performed on the light curves of different sunspot waves and 
oscillations with respect to the light curve of umbral flash location observed from AIA 1700~{\AA}. Cross-correlation coefficients
are plotted for different  time lags. Locations of the light curves are shown in Figure~\ref{fig:xt}.
Type of sunspot waves and oscillations considered for cross-correlation and corresponding passbands are labeled in the respective panels}.
\label{fig:correlaion}
\end{figure*}

We further refine our findings by obtaining oscillation amplitudes of different wave types shown in top panels of Figures~\ref{fig:ufwavelet},
\ref{fig:uwwavelet}, and \ref{fig:cwwavelet} and plotted in Figure~\ref{fig:growing}. Oscillation amplitudes are obtained with respect to 
the background signals, which were obtained from 8-min running average of original light curves as previously. Figure~\ref{fig:growing} clearly reveals
a similar pattern of growth in all the oscillation amplitudes. The amplitude of oscillations grew by more than 20\%
for umbral flashes observed in AIA 1600 \AA , whereas that for umbral and coronal waves grew up to $\approx 10\%$ and 5\% respectively. 
We also see a saw-tooth pattern where the amplitude first increases sharply, and later decreases slowly 
for umbral flash oscillations. This pattern is also visible in umbral and coronal wave amplitudes, however, to a lesser extent.
The appearance of the saw-tooth pattern may indicate the propagation of shock waves as suggested by \citet{2014ApJ...786..137T} in the
transition region lines. Similarity in the growing amplitude of oscillation, and almost co-temporal appearance of umbral flashes with 
those of umbral and coronal waves is a strong indication that these waves are influenced by umbral flashes.

To quantify the amplitude growth of these 2.8-min oscillations, we look at the oscillatory power of these waves with time.
Since the wavelet transform provides a temporally variable oscillatory power, we obtain the oscillatory power of these waves with time
using the wavelet transforms shown in Figures~\ref{fig:ufwavelet}, \ref{fig:uwwavelet}, and \ref{fig:cwwavelet}. 
Henceforth, we obtained the wavelet oscillatory power at around 2.8-min period averaged over the range of 2.3--3.3 min.
In Figure~\ref{fig:power}, we show oscillatory wavelet power for different oscillation and waves. The upper two panels 
are shown for coronal waves in AIA 171 and 211 \AA , middle panels for umbral waves in AIA 1600 and 304 \AA , and bottom panels 
for umbral flashes in AIA 1600 and 1700 \AA . On each  panel, we over-plot green curves to show the errors
associated with these oscillatory power curves. These error-bars are obtained by carrying out the same wavelet analysis on Monte Carlo 
bootstrapped light curves. In this method, we generate new light curves from the observed one, including point-wise error estimates on the intensities.
This is obtained by adding the normalized random distribution of errors to the original light curves. For the purpose, we generated 100 such new light curves.
Then we performed the same wavelet analysis to get a measure of the fuzziness in the results due to statistical fluctuations. 
Respective errorbars on AIA light curves were obtained using routine $ aia\_bp\_estimate\_error$ \citep{2012SoPh..275...41B}. 
The plots show almost similar power characteristics for all the waves. 
Given the range of errorbars, we conclude that consistent growth observed in wavelet powers (in the period range 2.3--3.3 min)
between 09:44:00 to 10:00:00 UT is real. Thus, findings of almost co-temporal increase of oscillatory power in around 2.8-min period further
strengthens our claim of association between umbral flashes and waves, and coronal waves.

\subsection{Time-Delay Analysis}
To further strengthen and understand the probable coupling among different waves and oscillations, we performed a cross-correlation 
analysis of these waves for the duration 09:43:00 UT to 10:00:00 UT. The time is chosen such that it covers the time of occurrence 
of the umbral flashes. This enables us to observe the time lags associated with the maximum correlation 
co-efficients, and hence, to determine the time delays between different waves. We choose the light curve of umbral flashes 
obtained using 1700~{\AA} images to perform the cross-correlation with light curves of 
umbral flashes observed in 1600~{\AA} and 304~{\AA}, umbral waves observed in 1600~{\AA}, and 304~{\AA}, and coronal 
waves observed in 171~{\AA}, and 211~{\AA}. 

Figure~\ref{fig:correlaion} displays the results of cross-correlation analysis in terms of correlation coefficient obtained for different 
time lags. The analysis is performed using standard IDL routine $c\_correlate$ that finds the correlations amongst the amplitude of 
oscillations of different sunspot waves and oscillations. Plots reveal around 70$\%$ correlation for all the waves with respect to AIA 1700~{\AA} 
umbral flash oscillations. We observe  an increase in time delay corresponding to the peak correlation coefficient as we go from chromospheric 
umbral flashes and umbral waves to coronal waves. The time delay increases because distance at which light curves were obtained increases 
for umbral waves and coronal waves with respect to umbral flash location (see Figure~\ref{fig:xt}). However, time delays obtained from AIA 304 \AA\ 
passband are relatively larger for umbral flash and wave as compared to AIA 1600 \AA\  passband. This may indicate that AIA 304 \AA\  forms at 
higher atmospheric height compared to  AIA 1700, and 1600 \AA\ passbands. Furthermore, we do not find any significant time delays 
among the coronal passbands. This could be attributed to the fact that emissions in different AIA passbands are coming from the lower 
temperature components as fanloops are typically of 1 MK temperature \citep[e.g.,][]{2017ApJ...835..244G}.
The significantly correlated light curves observed in chromospheric umbral flashes with umbral waves, and coronal waves, confirm 
the influence of umbral flashes on umbral waves and coronal waves.

\section{Summary and Conclusions} \label{conclusion}

In this paper, we have focused on different types of sunspot oscillations and waves observed at solar chromospheric and coronal heights. 
We explored the sunspot with AIA 1700~{\AA}, 1600~{\AA} and 304~{\AA} passbands and the fanloop region over it 
with AIA 131~{\AA}, 171~{\AA}, 193~{\AA}, 211~{\AA}, and 335~{\AA} passbands. We list our findings below:

\begin{enumerate}

\item Five bright umbral flashes were identified from AIA 1700~{\AA}, 1600~{\AA} 
and 304~{\AA} images (shown in Figures~\ref{fig:uf}, and\ ~\ref{fig:slit}). Their locations were found in close proximity 
to the footpoint of one of the fanloops that were rooted in the umbra (shown in Figure~\ref{fig:slit}).

\item Emergence of umbral waves moving radially outward was observed in AIA 304~{\AA}, 1600~{\AA}, and 1700~{\AA} 
passbands from the locations of umbral flashes (shown in Figure~\ref{fig:xt}). 
The amplitude of umbral waves increased during the umbral flashes.

\item Almost all the AIA coronal passbands showed signatures of propagating magneto-acoustic 
waves along the different fanloop structures of umbral and penumbral origin.
However, the fanloop systems that was rooted inside the sunspot umbra showed oscillations with modulations in amplitude (shown in Figure~\ref{fig:xt}).
Combined time-distance plot of chromospheric AIA 1600~{\AA}, and coronal AIA 171~{\AA} showed a simultaneous amplitude increase
in coronal waves that could be associated with the umbral flashes, and thus, with umbral waves (shown in Figure~\ref{fig:mxt}).
Hence, the increasing amplitude of the
coronal waves could be influenced by the occurrence of umbral flashes. Moreover, the umbral flash light curves, and sometimes
(to a lesser extent) umbral waves and coronal waves light curves reveal a clear saw-tooth pattern of oscillations (shown in Figure~\ref{fig:growing}),
which can be attributed to chromospheric response to the magneto-acoustic shock due to propagating photospheric p-mode oscillations
\citep[e.g.,][]{2006ApJ...640.1153C,2014ApJ...786..137T}.
 
\item Using wavelet analysis, we obtained periods of oscillation of the different sunspot waves. For all the waves, i.e., umbral flash, 
umbral waves, and coronal waves, the dominant period was $\approx$ 2.8-min (shown in Figure~\ref{fig:ufwavelet},~\ref{fig:uwwavelet},
and \ref{fig:cwwavelet}). The co-temporal growth of 2.8-min oscillations for all the sunspot waves and oscillations 
were also suggested by the temporal variation of wavelet power (shown in Figure~\ref{fig:power}, which shows 
simultaneous growth in wavelet power for all the sunspot waves and oscillations).

\item The significant correlations among chromospheric umbral flash, umbral waves, and coronal waves with some time delays
is an indication of propagation of sunspot oscillation and waves from the lower atmosphere to the upper atmosphere (shown in 
Figure~\ref{fig:correlaion}).

\end{enumerate}

The results obtained here provide the first direct observational evidence of the influence of chromospheric umbral flashes on umbral waves
and coronal waves. These results are supported by the time-distance maps and simultaneous growth in oscillation amplitudes obtained
from the original light curves. Though our results are based on the analysis of original, unfiltered light curves, we also performed the same analysis 
using the Fourier filtered light curves obtained within the frequency range 5--7 mHz ($\approx 2.3-3.3$ min). The Fourier filtered light curves also yielded
similar co-temporal pattern for different sunspot oscillation and waves in the different AIA passbands.
Our results point towards the occurrence of a few strong umbral flashes which influence the propagation of all sunspot waves and oscillations
observed at different solar atmospheric layers. Hence, we show the effect of chromospheric umbral flashes in the corona. 
The analysis presented here also provides important findings to understand trigger mechanism of coronal jets. \citet{2015MNRAS.446.3741C} 
suggested that jets were triggered due to increase in the amplitude of waves. This analysis provides the reason for the increase and therefore,
important results for initiations of jets. 

To further confirm and establish these findings, co-ordinated observations of sunspots waves and oscillations using simultaneous  ground
and space-based facilities are essential. The Solar Ultraviolet Imaging Telescope  \citep[SUIT;][]{2016SPIE.9905E..03G} on board Aditya-L1 
will provide excellent coverage of photosphere and chromosphere to study the coupling of these waves in more details.

\acknowledgments
AS thanks IUCAA for providing local hospitality and support during her stay.
GRG is supported through the INSPIRE Faculty Award of the Department of Science and Technology (DST), India. 
DT acknowledges the support from the Max-Planck Partner Group on Coupling and Dynamics of the Solar Atmosphere
at IUCAA. VK was supported through NASA Contract NAS8- 03060 to the Chandra X-ray Centre. Part of this work
was done during collaborative visits that were part of the ClassACT - an Indo-US Centre for Astronomical
Object and Feature Characterization and Classification, sponsored by the Indo-US Science and Technology
Forum (IUSSTF). AP acknowledges visiting associateship of IUCAA. Authors thank Drs. Dipankar banerjee,
David Jess, and Krishna Prasad for helpful discussions. AIA and HMI data are courtesy of SDO (NASA). 
Facilities: SDO (AIA, HMI).


\begin{thebibliography}{}
\expandafter\ifx\csname natexlab\endcsname\relax\def\natexlab#1{#1}\fi

\bibitem[{{Beckers} \& {Tallant}(1969)}]{1969SoPh....7..351B}
{Beckers}, J.~M., \& {Tallant}, P.~E. 1969, \solphys, 7, 351

\bibitem[{{Bloomfield} {et~al.}(2007){Bloomfield}, {Lagg}, \&
  {Solanki}}]{2007ApJ...671.1005B}
{Bloomfield}, D.~S., {Lagg}, A., \& {Solanki}, S.~K. 2007, \apj, 671, 1005

\bibitem[{{Boerner} {et~al.}(2012){Boerner}, {Edwards}, {Lemen}, {Rausch},
  {Schrijver}, {Shine}, {Shing}, {Stern}, {Tarbell}, {Title}, {Wolfson},
  {Soufli}, {Spiller}, {Gullikson}, {McKenzie}, {Windt}, {Golub}, {Podgorski},
  {Testa}, \& {Weber}}]{2012SoPh..275...41B}
{Boerner}, P., {Edwards}, C., {Lemen}, J., {et~al.} 2012, \solphys, 275, 41

\bibitem[{{Bogdan} \& {Judge}(2006)}]{2006RSPTA.364..313B}
{Bogdan}, T.~J., \& {Judge}, P.~G. 2006, Philosophical Transactions of the
  Royal Society of London Series A, 364, 313

\bibitem[{{Centeno} {et~al.}(2006){Centeno}, {Collados}, \& {Trujillo
  Bueno}}]{2006ApJ...640.1153C}
{Centeno}, R., {Collados}, M., \& {Trujillo Bueno}, J. 2006, \apj, 640, 1153

\bibitem[{{Chandra} {et~al.}(2015){Chandra}, {Gupta}, {Mulay}, \&
  {Tripathi}}]{2015MNRAS.446.3741C}
{Chandra}, R., {Gupta}, G.~R., {Mulay}, S., \& {Tripathi}, D. 2015, \mnras,
  446, 3741

\bibitem[{{Christopoulou} {et~al.}(2000){Christopoulou}, {Georgakilas}, \&
  {Koutchmy}}]{2000A&A...354..305C}
{Christopoulou}, E.~B., {Georgakilas}, A.~A., \& {Koutchmy}, S. 2000, \aap,
  354, 305

\bibitem[{{Christopoulou} {et~al.}(2001){Christopoulou}, {Georgakilas}, \&
  {Koutchmy}}]{2001A&A...375..617C}
---. 2001, \aap, 375, 617

\bibitem[{{De Moortel}(2009)}]{2009SSRv..149...65D}
{De Moortel}, I. 2009, \ssr, 149, 65

\bibitem[{{De Moortel} {et~al.}(2002){De Moortel}, {Ireland}, {Hood}, \&
  {Walsh}}]{2002A&A...387L..13D}
{De Moortel}, I., {Ireland}, J., {Hood}, A.~W., \& {Walsh}, R.~W. 2002, \aap,
  387, L13

\bibitem[{{De Moortel} \& {Nakariakov}(2012)}]{2012RSPTA.370.3193D}
{De Moortel}, I., \& {Nakariakov}, V.~M. 2012, Royal Society of London
  Philosophical Transactions Series A, 370, 3193

\bibitem[{{Ghosh} {et~al.}(2017){Ghosh}, {Tripathi}, {Gupta}, {Polito},
  {Mason}, \& {Solanki}}]{2017ApJ...835..244G}
{Ghosh}, A., {Tripathi}, D., {Gupta}, G.~R., {et~al.} 2017, \apj, 835, 244

\bibitem[{{Ghosh} {et~al.}(2016){Ghosh}, {Chatterjee}, {Khan}, {Tripathi},
  {Ramaprakash}, {Banerjee}, {Chordia}, {Gandorfer}, {Krivova}, {Nandy},
  {Rajarshi}, {Solanki}, \& {Sriram}}]{2016SPIE.9905E..03G}
{Ghosh}, A., {Chatterjee}, S., {Khan}, A.~R., {et~al.} 2016, in \procspie, Vol.
  9905, Society of Photo-Optical Instrumentation Engineers (SPIE) Conference
  Series, 990503

\bibitem[{{Gupta} {et~al.}(2010){Gupta}, {Banerjee}, {Teriaca}, {Imada}, \&
  {Solanki}}]{2010ApJ...718...11G}
{Gupta}, G.~R., {Banerjee}, D., {Teriaca}, L., {Imada}, S., \& {Solanki}, S.
  2010, \apj, 718, 11

\bibitem[{{Gupta} {et~al.}(2012){Gupta}, {Teriaca}, {Marsch}, {Solanki}, \&
  {Banerjee}}]{2012A&A...546A..93G}
{Gupta}, G.~R., {Teriaca}, L., {Marsch}, E., {Solanki}, S.~K., \& {Banerjee},
  D. 2012, \aap, 546, A93

\bibitem[{{Jess} {et~al.}(2012){Jess}, {De Moortel}, {Mathioudakis},
  {Christian}, {Reardon}, {Keys}, \& {Keenan}}]{2012ApJ...757..160J}
{Jess}, D.~B., {De Moortel}, I., {Mathioudakis}, M., {et~al.} 2012, \apj, 757,
  160

\bibitem[{{Jess} {et~al.}(2013){Jess}, {Reznikova}, {Van Doorsselaere}, {Keys},
  \& {Mackay}}]{2013ApJ...779..168J}
{Jess}, D.~B., {Reznikova}, V.~E., {Van Doorsselaere}, T., {Keys}, P.~H., \&
  {Mackay}, D.~H. 2013, \apj, 779, 168

\bibitem[{{Kiddie} {et~al.}(2012){Kiddie}, {De Moortel}, {Del Zanna},
  {McIntosh}, \& {Whittaker}}]{2012SoPh..279..427K}
{Kiddie}, G., {De Moortel}, I., {Del Zanna}, G., {McIntosh}, S.~W., \&
  {Whittaker}, I. 2012, \solphys, 279, 427

\bibitem[{{Kobanov} {et~al.}(2006){Kobanov}, {Kolobov}, \&
  {Makarchik}}]{2006SoPh..238..231K}
{Kobanov}, N.~I., {Kolobov}, D.~Y., \& {Makarchik}, D.~V. 2006, \solphys, 238,
  231

\bibitem[{{Kobanov} \& {Makarchik}(2004)}]{2004A&A...424..671K}
{Kobanov}, N.~I., \& {Makarchik}, D.~V. 2004, \aap, 424, 671

\bibitem[{{Krishna Prasad} {et~al.}(2011){Krishna Prasad}, {Banerjee}, \&
  {Gupta}}]{2011A&A...528L...4K}
{Krishna Prasad}, S., {Banerjee}, D., \& {Gupta}, G.~R. 2011, \aap, 528, L4

\bibitem[{{Krishna Prasad} {et~al.}(2015){Krishna Prasad}, {Jess}, \&
  {Khomenko}}]{2015ApJ...812L..15K}
{Krishna Prasad}, S., {Jess}, D.~B., \& {Khomenko}, E. 2015, \apjl, 812, L15

\bibitem[{{Lemen} {et~al.}(2012){Lemen}, {Title}, {Akin}, {Boerner}, {Chou},
  {Drake}, {Duncan}, {Edwards}, {Friedlaender}, {Heyman}, {Hurlburt}, {Katz},
  {Kushner}, {Levay}, {Lindgren}, {Mathur}, {McFeaters}, {Mitchell}, {Rehse},
  {Schrijver}, {Springer}, {Stern}, {Tarbell}, {Wuelser}, {Wolfson}, {Yanari},
  {Bookbinder}, {Cheimets}, {Caldwell}, {Deluca}, {Gates}, {Golub}, {Park},
  {Podgorski}, {Bush}, {Scherrer}, {Gummin}, {Smith}, {Auker}, {Jerram},
  {Pool}, {Soufli}, {Windt}, {Beardsley}, {Clapp}, {Lang}, \&
  {Waltham}}]{2012SoPh..275...17L}
{Lemen}, J.~R., {Title}, A.~M., {Akin}, D.~J., {et~al.} 2012, \solphys, 275, 17

\bibitem[{{Madsen} {et~al.}(2015){Madsen}, {Tian}, \&
  {DeLuca}}]{2015ApJ...800..129M}
{Madsen}, C.~A., {Tian}, H., \& {DeLuca}, E.~E. 2015, \apj, 800, 129

\bibitem[{{Pesnell} {et~al.}(2012){Pesnell}, {Thompson}, \&
  {Chamberlin}}]{2012SoPh..275....3P}
{Pesnell}, W.~D., {Thompson}, B.~J., \& {Chamberlin}, P.~C. 2012, \solphys,
  275, 3

\bibitem[{{Rouppe van der Voort} {et~al.}(2003){Rouppe van der Voort},
  {Rutten}, {S{\"u}tterlin}, {Sloover}, \& {Krijger}}]{2003A&A...403..277R}
{Rouppe van der Voort}, L.~H.~M., {Rutten}, R.~J., {S{\"u}tterlin}, P.,
  {Sloover}, P.~J., \& {Krijger}, J.~M. 2003, \aap, 403, 277

\bibitem[{{Sych}(2016)}]{2016GMS...216..467S}
{Sych}, R. 2016, Washington DC American Geophysical Union Geophysical Monograph
  Series, 216, 467

\bibitem[{{Thomas} \& {Weiss}(2008)}]{2008sust.book.....T}
{Thomas}, J.~H., \& {Weiss}, N.~O. 2008, {Sunspots and Starspots} (Cambridge
  University Press)

\bibitem[{{Tian} {et~al.}(2014){Tian}, {DeLuca}, {Reeves}, {McKillop}, {De
  Pontieu}, {Mart{\'{\i}}nez-Sykora}, {Carlsson}, {Hansteen}, {Kleint},
  {Cheung}, {Golub}, {Saar}, {Testa}, {Weber}, {Lemen}, {Title}, {Boerner},
  {Hurlburt}, {Tarbell}, {Wuelser}, {Kankelborg}, {Jaeggli}, \&
  {McIntosh}}]{2014ApJ...786..137T}
{Tian}, H., {DeLuca}, E., {Reeves}, K.~K., {et~al.} 2014, \apj, 786, 137

\bibitem[{{Torrence} \& {Compo}(1998)}]{1998BAMS...79...61T}
{Torrence}, C., \& {Compo}, G.~P. 1998, Bull. Am. Meteorol. Soc., 79, 61

\bibitem[{{Tziotziou} {et~al.}(2006){Tziotziou}, {Tsiropoula}, {Mein}, \&
  {Mein}}]{2006A&A...456..689T}
{Tziotziou}, K., {Tsiropoula}, G., {Mein}, N., \& {Mein}, P. 2006, \aap, 456,
  689

\bibitem[{{Zhao} {et~al.}(2016){Zhao}, {Felipe}, {Chen}, \&
  {Khomenko}}]{2016ApJ...830L..17Z}
{Zhao}, J., {Felipe}, T., {Chen}, R., \& {Khomenko}, E. 2016, \apjl, 830, L17

\bibitem[{{Zirin} \& {Stein}(1972)}]{1972ApJ...178L..85Z}
{Zirin}, H., \& {Stein}, A. 1972, \apjl, 178, L85

\end{thebibliography}

\end{document}